\documentclass[twocolumn,pre,nofootinbib,floatfix,superscriptaddress]{revtex4-1}

\usepackage{bbm}
\usepackage{graphicx}
\usepackage{amsmath,amsfonts,amssymb}
\usepackage{multirow}
\usepackage{rotating}
\usepackage{url}

\begin{document}
%
\title{Detecting Friendship Within Dynamic Online Interaction Networks}

\author{Sears Merritt}
\email[]{sears.merritt@colorado.edu}
\author{Abigail Z. Jacobs}
\email[]{abigail.jacobs@colorado.edu}
\affiliation{Department of Computer Science, University of Colorado, Boulder, CO 80309}

\author{Winter Mason}
\email[]{m@winteram.com}
\affiliation{School of Technology Management, Stevens Institute of Technology, Hoboken, NJ 07030}

\author{Aaron Clauset}
\email[]{aaron.clauset@colorado.edu}
\affiliation{Department of Computer Science, University of Colorado, Boulder, CO 80309}
\affiliation{BioFrontiers Institute, University of Colorado, Boulder, CO 80303}
\affiliation{Santa Fe Institute, 1399 Hyde Park Rd., Santa Fe, NM 87501}

\begin{abstract}
In many complex social systems, the timing and frequency of interactions between individuals are observable but friendship ties are hidden. 
Recovering these hidden ties, particularly for casual users who are relatively less active, would enable a wide variety of friendship-aware applications in domains where labeled data are often unavailable, including online advertising and national security. Here, we investigate the accuracy of multiple statistical features, based either purely on temporal interaction patterns or on the cooperative nature of the interactions, for automatically extracting latent social ties. Using self-reported friendship and non-friendship labels derived from an anonymous online survey, we learn highly accurate predictors for recovering hidden friendships within a massive online data set encompassing 18 billion interactions among 17 million individuals of the popular online game \textit{Halo: Reach}. We find that the accuracy of many features improves as more data accumulates, and cooperative features are generally reliable. However, periodicities in interaction time series are sufficient to correctly classify 95\% of ties, even for casual users. These results clarify the nature of friendship in online social environments and suggest new opportunities and new privacy concerns for friendship-aware applications that do not require the disclosure of private friendship information.
\end{abstract}

\maketitle

\section{Introduction}
For many online social systems, understanding which users are ``friends,'' can be extremely useful, e.g., for targeted word-of-mouth advertising, product recommendations, or detecting hidden social relationships.  In some systems these relationships are provided by the users themselves, but even when the friendships are not explicitly labeled, we can often still observe the timing and character of pairwise social interactions; for example, citations between scientists~\cite{price1965statistical}, appearances together in photos~\cite{crandall2010inferring}, exchanges of tweets~\cite{wu2011says}, emails~\cite{de2010inferring} or phone calls, playing games together, purchasing goods or services from businesses, etc.

This raises the question of whether hidden or latent friendship ties can be inferred from such interaction data alone. For most online systems, this is complicated by the typically heavy-tailed distribution in the volume of interactions generated by different users: only a small fraction of users account for the majority of all interactions, providing deep histories from which to learn, while most users are ``casual,'' generating relatively little data. Inferring latent ties from observable interactions promises to create both new opportunities and raise new privacy concerns for friendship-aware applications, e.g., in online advertising, where latent tie inference could facilitate social marketing or better estimate product preferences, and online security, where it could uncover clandestine associations and activities.

For many computational social science questions, online multiplayer games are a rich but underutilized source of detailed, temporal interaction data. Past work in this area has shed light on competitive dynamics, social organization, economic trading networks, and deviant behavior~\cite{szell2010multirelational,keegan2010dark,blackburn2012branded}. Here we utilize a massive data set from the popular online multiplayer game \textit{Halo: Reach} to investigate the degree to which latent social ties can be automatically identified from social interaction data alone. This data set contains details on more than 18 billion interactions among more than 17 million unique individuals across 700 million game instances, and serves as a model system by which to investigate the general question of detecting friendship in dynamic online interaction networks.

From these data, we extract a temporal interaction network, in which two individuals are connected at time $t$ if they shared a social interaction at time $t$. Here, interactions are playing a game together. We annotated each interaction with information about its character and magnitude, e.g., if it was a prosocial or antisocial interaction. We then combine these data with the results of an anonymous online survey of the player population~\cite{mason2012friends}, including friendship and non-friendship labels for every individual in their time series. 

We then design and study nine statistical features representing temporal and cooperative-type interactions. Temporal features capture interaction patterns via periodicities, interaction volume, and the similarity in actions within the online system. Cooperative features quantify the prosocial character of the interactions such as direct and indirect assistance in scoring points, and ``betrayals,'' the equivalent of scoring on one's own goal in the game, which indicates antisocial behavior toward the betrayed individual. Although our cooperative features rely on in-game data specific to \textit{Reach}, the intention here is to capture the character or sign of the interaction~\cite{leskovec2010predicting}, and thus analogous features can likely be constructed for other types of interaction data.  For instance, the interaction patterns in the game setting could correspond to check-ins with a location-based application; the cooperative features in the game could correspond to positive or negative comments on an online forum.

From a social theory perspective, temporal features are expected to provide a weaker signal than cooperative ones because the former ignore the additional information explicitly contained in the latter. On the other hand, temporal features are more generalizable because they can always be derived from interaction time series, even when auxiliary information is unavailable, e.g., to study co-location, online social interaction, and communication data \cite{cranshaw2010cmuisr,clauset2007persistence,eagle2009inferring,de2010inferring}.  In contrast to many standard data sets, our data allow us to directly compare the predictive utility of these two types of features.

The self-reported friend and non-friend labels from the online survey allow us to quantitatively measure the accuracy of our latent tie inference methods, and we take a supervised approach to learn which features perform well at this task.  We also explore the way their performance degrades as we examine ties with progressively less data, which is an important concern for real-world applications. In general, we find that latent friendship ties can be predicted with over 95\% accuracy when two individuals have had at least 10 interactions. This level of accuracy is achievable using either the auto-correlation of interaction (temporal) or the number of assists (cooperation). The total volume of interactions between individuals is also a good predictor, but it is less efficient than our two best features. These results clarify the nature of friendship in online social environments and suggest new opportunities and new privacy concerns for friendship-aware applications that do not require the disclosure of private friendship information.

\section{Related work}
\label{sec:related:work}
Our work draws from three distinct lines of research. Most uses of online game data have focused on understanding certain aspects of human social behavior in online environments. Examples include individual and team performance~\cite{shim2010player,shim2010team,shim:etal:2011,shim2011exploratory}, expert behavior~\cite{huffaker2009social}, homophily~\cite{huang2009virtually}, group formation~\cite{huang2009formation}, economic activity~\cite{castronova2009real,bakshy2010social}, and deviant behavior~\cite{ahmad2009mining}. Most of this work has focused on massively multiplayer online role playing games (MMORPGs), e.g., World of Warcraft, although a few have examined social behavior in first person shooter (FPS) games like \textit{Reach}~\cite{shim:etal:2011}. Relatively little of this work has focused on the structure of social networks. 

Some studies in social network analysis have considered human behavioral patterns in proximity and periodicity, e.g., questions regarding how the accumulation of interactions over time or physical proximity and geographic location can influence the induced social network structure~\cite{clauset2007persistence,eagle2009inferring,de2010inferring,crandall2010inferring}. Few of these studies have focused on online interactions and the way they reflect underlying social ties.

Another significant thread comes from the literature on link prediction. Several studies have considered the question of predicting links in future time steps based on the pattern of links in the past~\cite{liben2007link}. Others have focused on predicting hidden or missing links when given a partially observed network~\cite{clauset2008hierarchical,sarkar2012nonparametric}, and on how similarities in preferences and periodic behavior can predict social ties and their sign (friend or foe, trust or distrust)~\cite{adamic2003friends,li2008tag,eagle2009inferring,crandall2010inferring,leskovec2010predicting}. 

Of particular relevance is a recent study that applied a similar approach to ours, with good results,  to the more narrow question of distinguishing close and not close friends among a user's ties on Facebook~\cite{jones2013inferring}. Otherwise, very few studies have focused on the specific question and context considered here. A distinguishing feature of our study is the use of survey data, which provides us with ``ground truth'' labels of subjective friendship or non-friendship for observed interactions. By combining these ground-truth labels with the detailed data on pairwise social interactions among all individuals, we directly explore the question of distinguishing mere interactions from genuine latent friendships.

\section{Data and survey}
\label{sec:data:survey}
\subsection{Game details}
Our interaction data are drawn from \textit{Halo: Reach}, a popular online first person shooter game. It was publicly released by Bungie Inc., a former subdivision of Microsoft Game Studios, on 14 September 2010, and has generated more than 1 billion games since. Within the \textit{Reach} system, individuals choose from among seven game types and numerous subtypes, which are played over more than 33 terrain maps.
Games can be played alone or with or against other individuals over the Xbox Live online system, and each individual on the system is identified by a unique ``gamertag.'' Players may choose from among several ``playlists,'' which subdivide the total player population and which are based around specific game types. 

Once a playlist is chosen, individuals or small ``parties'' of players (typically friends) are grouped into teams by an in-game ``matchmaking'' algorithm. This algorithm is based on the TrueSkill system~\cite{herbrich2007trueskill}, which attempts to create teams with equal total skill (subject to some practical constraints). When a competition is complete, by default all its players are placed in a new game together, but all players or any subset may choose to reenter the matchmaking process to find new teammates or competitors. Both individual game and individual player summaries were made available through the Halo Reach Stats API.\footnote{The API was active from September 2010 through November 2012. API documentation was taken offline in September 2012.}

Through this interface, we collected the first 700 million game instances (roughly 305 days of activity by 17 million individuals). Among other information, each game file includes a Unix timestamp, game type label, and a list of gamertags. Each gamertag is associated with a particular team and a set of attributes indicating specific cooperative behavior actions amongst the individuals, described below. This large database provides us with complete data on the timing and character of interactions between individuals but provides no information about which interactions are produced by friendships versus non-friendships.

\subsection{Survey}
We combine these in-game behavioral data with the results of an anonymous online survey of \textit{Reach} players~\cite{mason2012friends}. In the survey, participants supplied their gamertag from which we generated a list of all other gamertags that had ever appeared in a game with the participant. From this list, the participant identified which individuals were friends.~\footnote{In the survey a friend is defined as a person known by the respondent at least casually, either offline or online.}  We interpret these subjective friendship labels as ground truth. From these data, we constructed a social network with links pointing from participants to their labeled friends. In our supervised learning analysis, both a labeled friendship and the absence of a label are treated as values to be predicted (i.e., we assume survey respondents explicitly chose not to label their co-player as a friend). Of the 965 participants who had completed the friendship portion of the survey by April 2012, 847 individuals appear in our data (the first 305 days of play); this yielded 14,045 latent friendship ties and 7,159,989 non-friendship ties.

Survey participants were a sparse sample of a large population, and the resulting social network is a composed of mostly disconnected egocentric subgraphs. Labeled friendship ties are directed edges, while observed interactions are bidirectional. We note that because survey participants were recruited through advertising on web fora related to \textit{Halo: Reach}, they are a non-uniform sample of the general \textit{Reach} population, e.g., they tended to be unusually skilled players~\cite{mason2012friends}.  Nonetheless, our sample has sufficient variability to demonstrate the general applicability of our results across the player population.

\subsection{Interaction network}
We represent the set of pairwise interactions as an annotated temporal network, in which edges have endpoints, exist at a specific moment in time, and are decorated with auxiliary information on the character and context of the interaction. Vertices in the network correspond to gamertags, and two vertices are connected if they appear in a game instance together at time $t$ (time of day, in 10 minute intervals). Each vertex thus has a sequence or time series of interactions with other vertices. We then annotate each edge with information like whether the corresponding individuals were on the same team, what game type produced the interaction, and number of games played together at time $t$. The resulting network, derived from our complete game sample, contains 17,286,270 vertices, 18,305,874,864 temporal edges, and spans $305$ days. The subgraph of interactions by our survey participants contained a total of 2,531,479 vertices and 665,401,283 temporal edges over the same period of time.

\section{Inferring friendship}
\label{sec:prediction}
To recover latent friendship ties given only the time series of annotated interactions between pairs of individuals, we take a supervised learning approach. Using classification trees and a logistic regression classifier~\cite{bishop2006pattern}, we learn which features are best for predicting latent friendship ties. Of particular interest will be computationally lightweight models that could be applied on large scale systems.

The self-reported friendship and non-friendship labels from the anonymous online survey serve as prediction targets. We investigate the accuracy of our statistical features, divided into temporal and cooperative classes and considered individually, for predicting latent ties. Temporal features are derived explicitly from a time series of interactions, without regard to the character or context of those interactions. Cooperative features are derived from the auxiliary data and capture the degree to which an interaction is prosocial. In the construction of several features, we use the massive unlabeled data to derive simple statistical expectations that are used to normalize the raw statistics.

\subsection{Temporal features}
\label{sec:general}

\begin{figure}[t!]
\begin{center}
\includegraphics[scale=0.42]{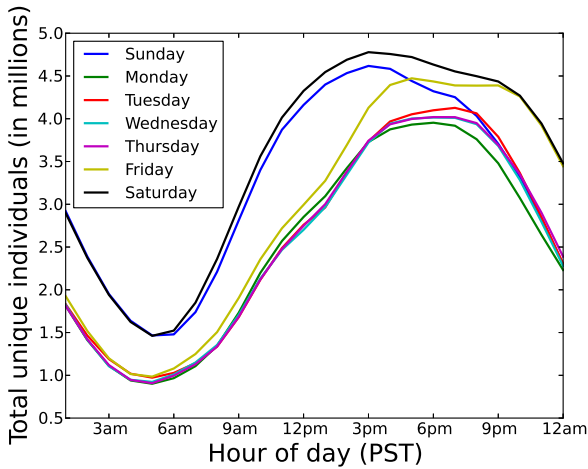}
\caption{Number of unique individuals ever seen at a given time of day (in Pacific Standard Time), across the 305 days spanned by the data, illustrating significant daily and weekly periodicities.}
\label{fig:players:per:hour}
\end{center}
\end{figure}

Overall gameplay dynamics within the \textit{Halo:\!\! Reach} system are highly periodic (Fig.~\ref{fig:players:per:hour}), with the peak online population on each day of the week occurring between the hours of 3:00pm and 6:00pm Pacific Standard Time (PST) and the minimum occurring near 4:30am. Since most players reside in the US and the majority of the US population is located on either the East or West coasts, the three hour window of peak play seems likely related to the coasts' three hour time difference. Furthermore, the peak period is roughly synchronized with the class schedules of secondary and post-secondary schools, where the majority of classes occur between the hours of 8:00am and 2:00pm. Finally, we observe a strong weekend effect, with Friday night game play rising to weekend levels, Saturday play remaining high and steady for the majority of the day and night, and Sunday play peaking relatively early and then tapering off after roughly 3:00pm. These regularities suggest several statistical features for capturing latent friendship ties.
\\

\noindent \textbf{Pair autocorrelation}. Pairs of individuals in \textit{Reach} that are friends are known to play many more consecutive games (12, on average, or about 2 hours of time) than non-friends (1.25, on average)~\cite{mason2012friends}. Thus, continuous interaction over a significant span of time is likely an indication of a latent tie, while more intermittent interactions likely indicate a non-friend tie, given the large population of non-friends available to play at any time. The expected diurnal and weekly cycles observed in the data will modulate these behaviors, and a reasonable approach for their quantification is via interaction periodicity. Let
\begin{align}
n_{x,y}(t) = \mathbbm{1}\{\textrm{$x$ and $y$ play together at time $t$}\}
\end{align}
represent the time series of binary interactions between individuals $x$ and $y$, where $1$ indicates an interaction at time $t$ and $0$ indicates no interaction. If $x$ and $y$ are friends, we expect $n_{x,y}(t)$ to exhibit stronger periodicity than for non-friends. This expectation may be quantified as the autocorrelation of the time series $n_{x,y}(t)$ over all time lags $\tau$:
\begin{align}
AC_{x,y} & = \sum_\tau \sum_t n_{x,y}(t)n_{x,y}(t-\tau).
\end{align}
If $n_{x,y}(t)$ is generated by a non-friend pair, $AC_{x,y}$ should be small because these individuals do not interact regularly. On the other hand, if $n_{x,y}(t)$ is generated by a friend pair, we expect $AC_{x,y}$ to be large. \\

\noindent \textbf{Pair frequency}. A corollary of our previous argument is that friend pairs will likely produce a greater number of interactions over a fixed time period than non-friend pairs. Let $N_{x}$ be the total number of games played by individual $x$, and
\begin{align}
N_{x,y} = \sum_t n_{x,y}(t)
\end{align}
be the number of those games played with individual $y$. The fraction $N_{x,y}/N_x$ thus captures the share of $x$'s interactions that involve $y$. Because we expect friend pairs produce more interactions than non-friend pairs, this fraction should be relatively large for a latent friend pair, even if the total number of $x$'s interactions, $N_{x}$, is small.\\

\noindent \textbf{Individual entropy}. Recent research has shown that individuals who maintain diverse or unpredictable patterns in their daily schedules in the physical world tend to have larger numbers of friends, as quantified by an entropy measure~\cite{cranshaw2010cmuisr}. But, online environments differ from physical ones in important ways, being more flexible and offering fewer constraints on ``large'' movements. It is thus an interesting question whether a digital version of these entropy measures can predict latent social ties as well as its physical analog.

Toward this end, we define entropy measures on an individual's schedule (when they interact), game type (in which game context do they interact), and combined schedule and game type. For a given individual $x$, we observe the series of $x$'s appearances at ``location'' $\ell \in \mathcal{L}$, where $\mathcal{L}$ represents the set of all possible locations. We consider three versions of this measure: (i) schedule entropy $H_{t}(x)$, with locations as days of the week, (ii) spatial entropy $H_{s}(x)$, with locations as \textit{Reach} ``playlists'' (which subdivide the full population into groups wanting to play a specific type of game), and (iii) the entropy $H_{s,t}(x)$ over all pairs of schedule and spatial locations.

Mathematically, we compute a given entropy measure as
\begin{align}
H_{\mathcal{L}}(x) = - \sum_{\ell\in\mathcal{L}}p(x,\ell)\log p(x,\ell),
\end{align}
where $p(x,\ell)$ corresponds to the observed probability of individual $x$ at location $\ell$, i.e., the fraction of all observations of $x$ in which $x$ is observed at location $\ell$. We expect the schedule entropy to quantify the diversity of an individual's interactions across time: individuals who typically play on Tuesdays (say, at 8:00pm to meet their friends) will have a lower entropy than those who play in more ad hoc fashions. Similarly, we expect the combined schedule-location entropy to capture regularities such as playing in one game environment on Tuesdays but in different environments over the rest of the week.  

For predicting friendships, we take the sum of the individuals' entropies, i.e., $H_{t}(x) + H_{t}(y)$, as opposed to a joint entropy measure. A low sum of entropy measures would suggest that both players have low diversity playing patterns, which need not be coordinated. A higher sum would suggest that at least one player of the pair has a more unpredictable schedule; however, knowing this is true for only one player is sufficient to suggest that other temporal signals might be more meaningful. An individual that plays sporadically but with a few regularities (e.g., consistently playing on Saturday mornings with the same set of individuals) suggests evidence of social coordination. A low entropy pair would then likely be either highly autocorrelated if they played on similar schedules, or exhibit very low autocorrelation if on different schedules. A rich class of temporal features lets us better describe the temporal patterns exhibited by the players in our sample and test existing hypotheses~\cite{cranshaw2010cmuisr}.

\subsection{Cooperative features}
\label{sec:domain:specific}
Our temporal features explicitly ignore the character of the interactions. Recent work and previous results suggest that friend pairs interact differently than non-friend pairs, and features that capture these differences can be expected to be good predictors of latent ties~\cite{hanaki2007cooperation,mason2012friends}.\\

\noindent \textbf{Betrayals}. One feature of \textit{Reach} that differs from many other online social systems is the ability to commit an explicitly antisocial action, in the form of a ``betrayal.'' These actions are equivalent to an ``own goal'' and result in a penalty for the betrayer's team.  A quirk of the method by which \textit{Reach} places players into a game is that occasionally friends are placed on opposing teams. Past work has shown that when this happens, one team tends to experience an increased betrayal rate as friends on one team turn against their teammates to help their friends on the other team~\cite{mason2012friends}.

For a pair of individuals $x$ and $y$, we capture this tendency by counting betrayals by $x$ that help $y$, i.e., when $x$ and $y$ are on different teams. Let $b_x(t)$ count the number of betrayals performed by $x$ at time $t$. Our measure is then
\begin{align}
B_{x,y} = \sum_{t} b_x(t) \,\mathbbm{1}\{ \textrm{$x$, $y$ playing on different teams} \}.
\end{align}
\noindent \textbf{Direct assistance}. During a game instance, individuals can provide direct assistance to each other in scoring a point. Like betrayals, this prosocial action can occur with or without deliberate coordination of actions. Because friend pairs are expected to exhibit greater frequencies of prosocial behavior toward each other, a large number of direct assists should correlate with latent friendship ties.

Let $a_x(t)$ count the number of direct assists performed by individual $x$ at time $t$. The total number of assists $A_{x,y}$ capture the volume of prosocial behavior on this tie,
\begin{align}
A_{x,y} = \sum_{t} a_x(t) \,\mathbbm{1}\{ \textrm{$x$, $y$ playing on same team} \}.
\end{align} 
\noindent \textbf{Indirect assistance}. \textit{Reach} also allows an individual to indirectly assist another in scoring points, in which $x$ drives a vehicle while $y$ operates a vehicle-mounted gun. This behavior requires substantially more coordination than direct assists, and thus may provide a more informative measure of latent friendship.

Let $v_x(t)$ count the number of indirect assists attributed to $x$ at time $t$. The total number of indirect assists from $x$ to $y$, denoted $V_{x,y}$, is 
\begin{align}
V_{x,y} = \sum_{t}  v_x(t) \,\mathbbm{1}\{ \textrm{$x$, $y$ playing on same team} \}.
\end{align}
%

\begin{figure}[t!]
\begin{center}
\includegraphics[scale=0.41]{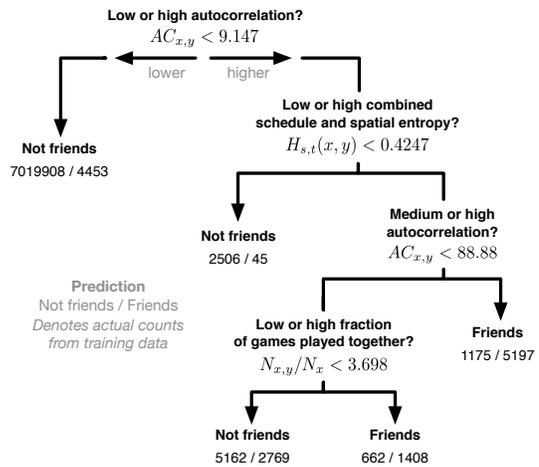}
\caption{A classification tree found using all features except $A_{x,y}$. This tree only uses temporal features, and performs well: the error rate is 0.0013, which is significantly better than the na\"ive classifier error rate of 0.0020. The out-of-sample AUC for this tree is 0.924.\textsuperscript{\ref{fnt:rnd:smp}}}
\label{fig:tree:allfeats}
\end{center}
\end{figure}

\subsection{Predicting latent friendships}
\label{sec:general:vs:specific}
In our initial exploration of the predictability of latent ties from interaction data, we use classification trees to gain intuition about which features or combinations thereof are likely to be predictive. For this data exploration, the interpretability of classification trees is a strength, compared to, e.g., random forests\footnote{To aid interpretation of the tree results, we normalize feature values by the average observed values taken from a uniform random sample of roughly 1 million players. For each of the players in the random sample we compute feature values for each player they interacted with in the data. \label{fnt:rnd:smp}}. Subsequently, we will consider the performance of individual features.

For learning the classification tree, we divided our data into equally sized groups of individuals for testing and training. Cross-validation within the test set was used to control the tree's complexity, pruning branches that did not significantly improve the fit of the model. The resulting tree is highly compact, with only a few features being retained (Fig.~\ref{fig:tree:allfeats}). Repeating our analysis with different subsets of the features and different training and test sets allows us to probe their relative importance and correlation structure.

All of the resulting trees beat the baseline accuracy of a na\"ive classifier. This baseline is in fact a significant barrier because the number of latent ties is a small fraction (0.2\%) of the total number of ties we consider and we can na\"ively score well by guessing that every tie is a non-friend.  For this reason, we use the Receiver Operating Characteristic (ROC) curve and the Area Under the ROC Curve (AUC)~\cite{bradley1997use}, which gives the probability the classifier will rank a randomly selected positive case higher than a randomly selected negative case.

At the level of feature classes, temporal features are most useful for correctly predicting friendship: when trained on all features, the best tree splits first on autocorrelation $AC_{x,y}$, followed by splits on combined schedule and spatial entropy $H_{s,t}(x,y)$, autocorrelation $AC_{x,y}$ (again), and normalized pair frequency $N_{x,y}/N_x$. Similar trees are found when training across all features excluding direct assists $A_{x,y}$, or only temporal features: for all three feature sets (all features, all features except assists, and temporal features only), the final trees yield average AUC scores of 0.830, 0.833, and 0.834 respectively. This similarity in performance is unsurprising considering the importance of temporal features (Fig.~\ref{fig:tree:allfeats}).

Surprisingly, fitting the model with just the cooperative features yields classification probabilities nearly as high (average AUC=$0.789$). This tree splits first on direct assistance $A_{x,y}$, in agreement with our expectation that latent friendship ties produce greater volumes of prosocial interactions than non-friend ties, followed by further splits on $A_{x,y}$ and indirect assistance $V_{x,y}$ over certain ranges of $A_{x,y}$. The fact that autocorrelation rather than direct assistance appears in the full model suggests first that autocorrelation is a more reliable indicator of latent friendship, but also that direct assistance may be capturing similar information. We test this idea by first training a classification tree using all features except autocorrelation $AC_{x,y}$. As expected, this tree splits first on high $A_{x,y}$, with the remaining structure being nearly identical to the models trained on all features or a subset, but substituting $A_{x,y}$ for $AC_{x,y}$. The average out-of-sample AUC for this set of trees is 0.800.

\begin{table*}[t]
\begin{center}
\begin{tabular}{c|lcrrrcc}
&  \multicolumn{1}{c}{feature} &  \multicolumn{1}{c} {$\gamma$} & \multicolumn{1}{c}{$\hat{\theta}$} & \multicolumn{1}{c}{$\hat{\sigma}$} & \multicolumn{1}{c}{$|Z|$} & $p$ & AUC \\ \hline
\multirow{6}{*}{\begin{sideways}temporal\end{sideways}} & pair autocorrelation & $AC_{x,y}$ & 0.0003 & 0.00001 & 30.000 & $\ll0.001$ & 0.99 \\
& normalized pair frequency & $N_{x,y}/N_x$ &  0.1390 & 0.00160 & 86.875 & $\ll0.001$ & 0.76 \\
& pair frequency & $N_{x,y}$ & 0.0390 & 0.00050 & 78.000 & $\ll0.001$ & 0.76 \\
& loc.\ entropy & $H_{s}(x)$ & 1.8270 & 0.04300& 42.488 & $\ll0.001$ & 0.65 \\
& sched.\ entropy & $H_{t}(x)$ & 1.5860 & 0.08100 & 19.580 & $\ll0.001$ & 0.50 \\
& sched.\ and loc.\ entropy& $H_{s,t}(x)$ & 2.5920 & 0.09600 & 27.000 & $\ll0.001$ & 0.61 \\ \hline 
\multirow{5}{*}{\begin{sideways}cooperative\end{sideways}} & & & & & \\
& direct assists & $A_{x,y}$ & 0.1230  & 0.00100 & 123.000 & $\ll0.001$ & 0.98 \\
& indirect assists & $V_{x,y}$ &  1.3170 & 0.01700 & 77.470 & $\ll0.001$ & 0.70 \\
& betrayals & $B_{x,y}$ & 0.1460 & 0.00300 & 48.590 & $\ll0.001$ & 0.64 \\
& & & & & \\
\end{tabular}
\end{center}
\caption{Coefficients, $\hat \theta$, standard deviations, $\hat \sigma$, Z-scores, $|Z|$, p values, $p$, and AUC values for logistic regression models fitted to each individual feature for all friends and non-friends. AUC values of 0.5 correspond to a baseline random classifier.}
\label{tab:all:features}
\end{table*}

The structure and simplicity of the fitted trees suggest an underlying signature of friendship in the patterns of observed interactions. Specifically, highly periodic interactions are strongly indicative of friendship because they require nontrivial levels of social coordination within the online environment. That is, friends must, and do, actively seek out each other in order to interact. Interestingly, although autocorrelation is highly predictive, combining it with spatial and schedule entropy reveals some subtleties in social interactions. When given all features or only temporal features, high autocorrelation $AC_{x,y}$ with high spatial and schedule entropy $H_{s,t}(x,y)$ yields a good predictor of latent friendships.\footnote{Note that while the classification tree only classifies friends and non-friends, the numbers observed, shown in the leaves of Figure \ref{fig:tree:allfeats}, indicate the maximum likelihood estimates of friendship probability at the leaf.}
Entropy features by themselves are not particularly useful, but they do become predictive for high values of autocorrelation. Players with shared, low diversity playing habits (and thus low individual entropy levels) can appear in the data as synchronized, even without any social coordination. Entropy measures then allow us to identify non-friends who have autocorrelated schedules.

\subsection{Lightweight predictors of friendship}
\label{sec:single:feature:perf}
These results suggest that individual features alone may perform well at predicting latent friendships, and such features would make good computationally lightweight predictors that could realistically be deployed on a large-scale system.

\begin{figure}[t!]
\begin{center}
\includegraphics[scale=0.41]{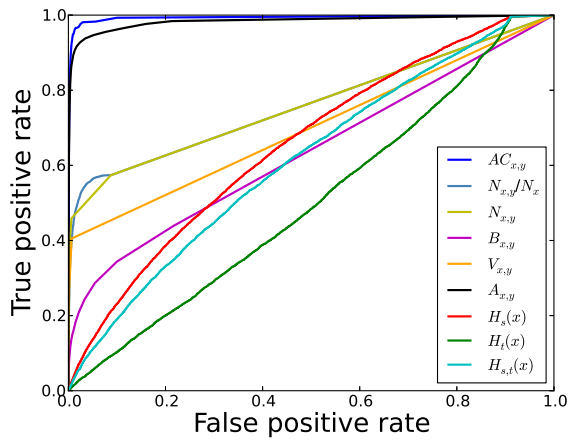}
\caption{ROC curves for logistic regression models on individual temporal and cooperative features.}
\label{fig:roc:all}
\end{center}
\end{figure}

We explore this possibility using logistic regression to build single-feature latent tie classifiers and measure their performance using AUC. We divide our data into training and test sets using random partitions such that test and training sets are of equal size.\textsuperscript{\ref{fnt:rnd:smp}} Figure~\ref{fig:roc:all} shows the ROC curves for each of these individual-feature models for predicting latent friendships, and the corresponding models are summarized in Table~\ref{tab:all:features}. Remarkably, the two most predictive individual features---autocorrelation $AC_{x,y}$ (temporal) and direct assistance $A_{x,y}$ (cooperative)---achieve near-perfect classifications, with AUCs of $0.99$ and $0.98$ respectively. To provide a comparison, we note that another method inferred friendship between graduate students with 96\% accuracy using a single temporal-spatial feature~\cite{eagle2009inferring}. Both of our single-feature models are computationally lightweight and could thus potentially be deployed on a large-scale system to automatically infer latent ties for friendship-aware applications.

\begin{figure*}[t!]
\begin{center}
\begin{tabular}{cc}
\includegraphics[scale=0.415]{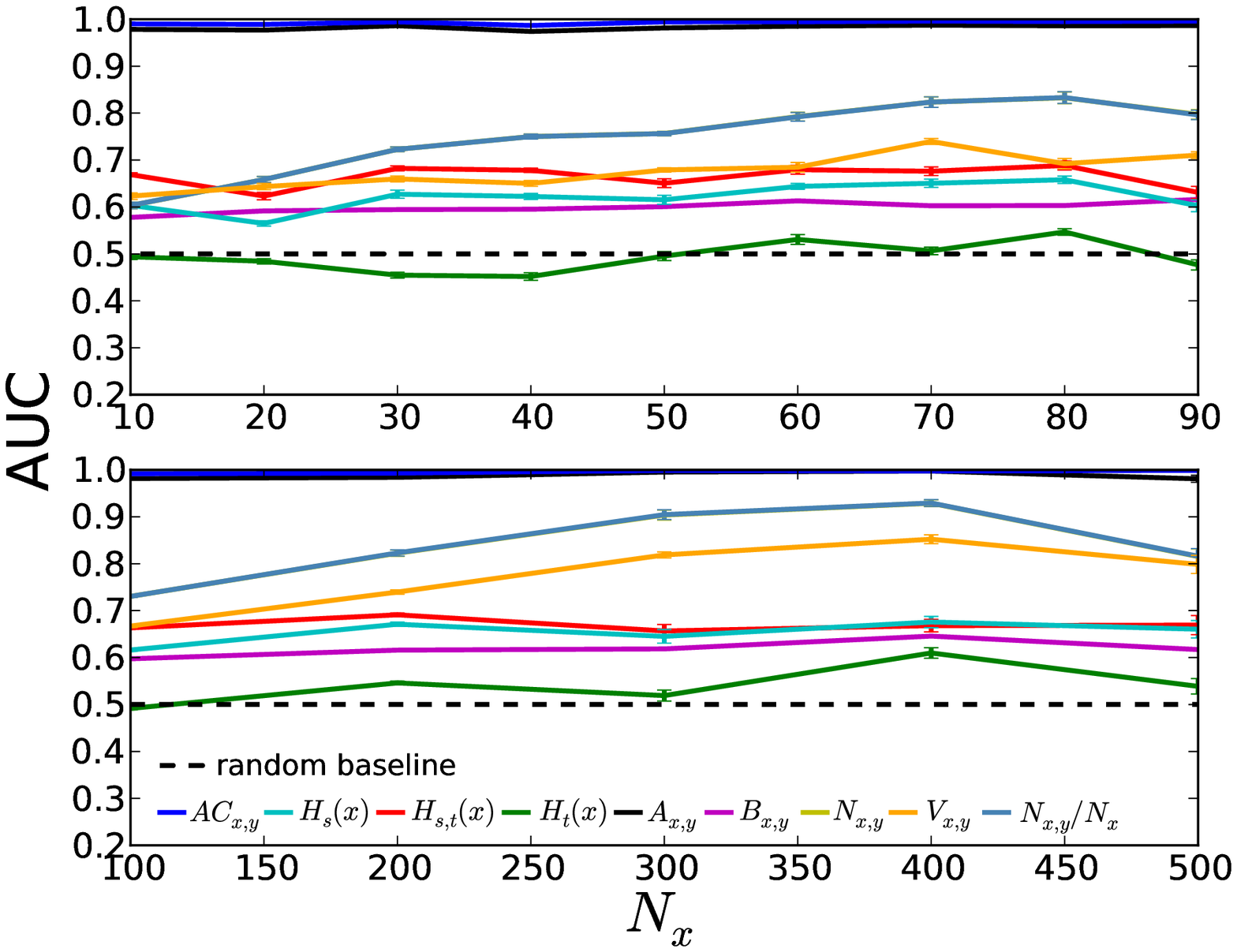} &
\includegraphics[scale=0.415]{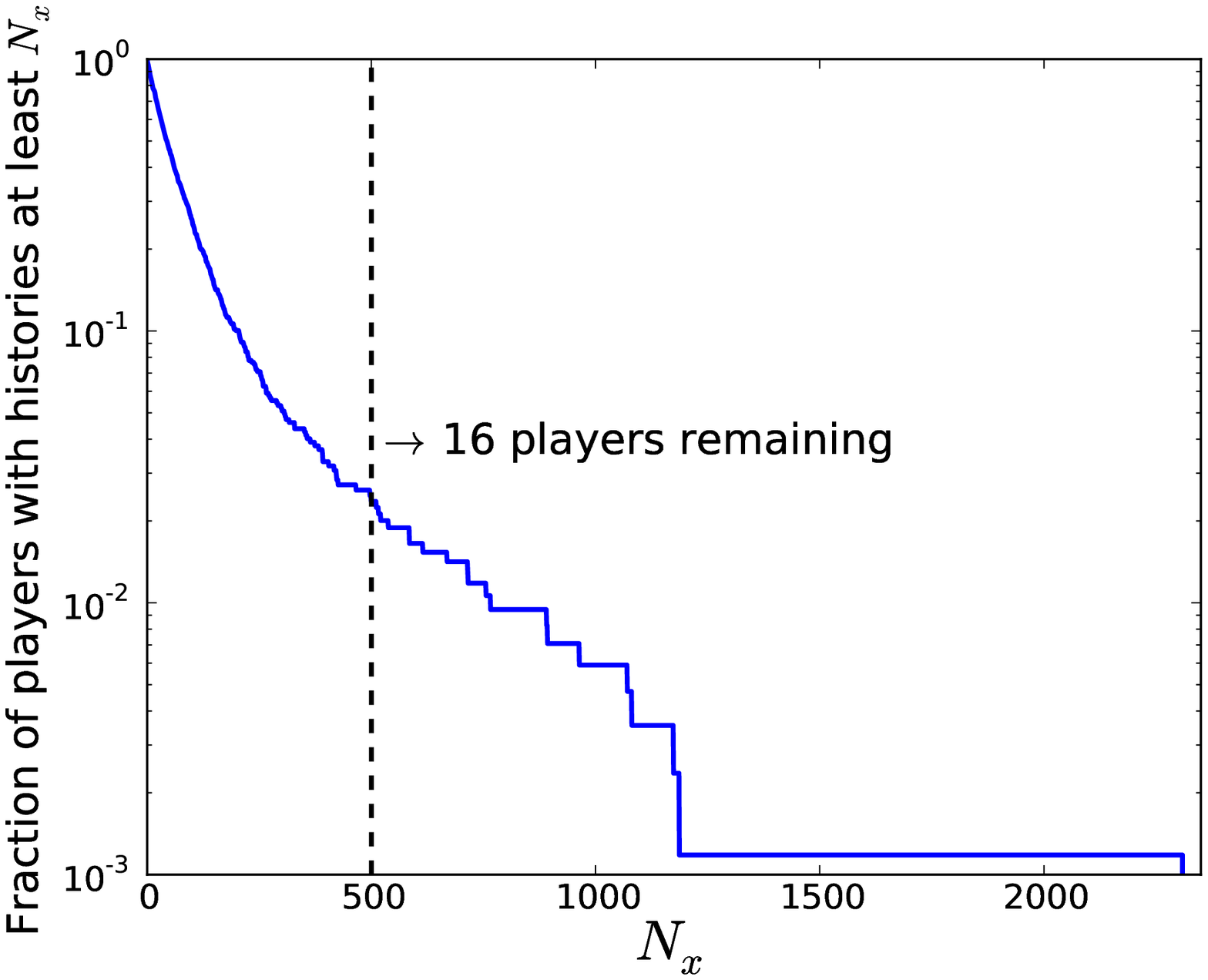}
\end{tabular}
\caption{(Left) AUC as a function of $N_x$ for each temporal and cooperative feature. The accuracy of $AC_{x,y}$ and $A_{x,y}$ are robust to available individual information while the accuracy of $V_{x,y}$, $N_{x,y}$ and $N_{x,y}/N_{x}$ increase with $N_x$. Entropic features remain relatively noisy regardless of $N_{x}$, see text for details. (Right) CCDF of $N_{x}$, number of games played, across all surveyed individuals.}
\label{fig:auc:ccdf}
\end{center}
\end{figure*}

All of the remaining individual features perform more poorly, indicating that none would perform well as lightweight predictors in a real-world environment. Na\"ively, we expected the volume of interaction $N_{x,y}$, and the fraction of that volume assigned to a particular other individual $N_{x,y}/N_{x}$, to be good indicators of latent ties. However, we find this not to be the case. Upon a closer examination of the mislabeled ties, we see that some latent ties spanned only a few interactions and this number was not significantly greater than the number of interactions with non-friends. Our autocorrelation feature is robust to this phenomenon because even these low-volume friendship ties exhibit strong periodicity in the interactions they generate.

Entropic features perform poorly alone because of insufficient diversity in location behavior within the population at large. That is, the number of interacting individuals at any given time is large, while the number of ``locations'' is relatively small. As a result, both friend and non-friend pairs will often make similar choices about which locations to visit. Controlling for both time and space via $H_{s,t}(x)$ provides a narrower filter to individuals' behavior but does not substantially improve performance. Furthermore, our entropy measure does not consider the alignment of the individuals' schedules. As we saw with the classification trees, it is only in combination with other features, like autocorrelation, that entropy becomes predictive.

The failure of entropy features alone to perform well in \textit{Reach} is interesting, and clarifies their success in applications to physical locations~\cite{cranshaw2010cmuisr}. When the number of locations is large relative to the size of the population exploring them, the probability becomes very low that a non-friend pair will have similar distributions over locations in time. As the number of locations shrinks relative to the population size, this probability increases and eventually swamps the signal produced by friend pairs, which is what we observe in \textit{Reach}. However, combining this signal with other features, like the autocorrelation, preserves some of its predictive power by mediating temporal effects with surprisingness, even in a system with densely occupied locations.

The poor performance of indirect assistance is unexpected, given that such behavior in \textit{Reach} indicates a strong prosocial orientation and that direct assistance performs so well. Examining the mislabeled ties, we find that indirect assistance is not always possible in every interaction, i.e., in every game type, and even when it is possible, it is an uncommon event. These factors place tight constraints on its predictive power and the raw behavioral data we study contain examples of labeled friend pairs that exhibit no indirect assistance, thus making it difficult to identify a discriminative threshold.

Past work on friendship in \textit{Reach}~\cite{mason2012friends} suggested that our betrayal feature (in which an individual betrays their teammates to help their friends on the opposite team) should also correlate with latent friendship. And indeed it does: the average betrayal total $\langle B_{x,y}\rangle = 6.27$ for friend pairs but only 0.5 for non-friend pairs. The significance of this difference is qualified by a substantially larger variance for friend pairs ($\sigma=29.12$ versus $2.13$), likely because many friends choose not to defect against their teammates, which lowers the discriminative power of this feature.

\subsection{Predicting friendships for casual users}
\label{sec:accuracy}
Achieving good predictions for the few users who produce large amounts of interaction data is useful. However, it is less useful if the performance degrades substantially as we consider users with progressively fewer observations, i.e., the casual users who typically make up the majority of individuals in an online system. To understand how robust our features are to the amount of available information, we study the performance of each individual feature as a function of $N_{x}$, the length of an individual's history. 

We grouped surveyed individuals into bins according to the number of games they completed $N_{x}$. To provide a fine-grained look at individuals with short histories, where data are plentiful, and a coarse view of long histories, where data are sparse (Fig.~\ref{fig:auc:ccdf}, right), we used bins of size 10 for $N_{x}<100$ and bins of size 100 for $N_{x}\ge100$.

We then computed the average AUC and its standard error by creating equal sized training and test sets from 10 random permutations of the data in each bin, and applying the individual-feature models. Examining these predictors' performance as a function of data volume provides some guidance for predicting friendships in data sets with large heterogeneities in data availability. Additionally, this test serves as a robustness check on our previous conclusions by implicitly considering the length of individual history as a feature.

Figure~\ref{fig:auc:ccdf} shows the average AUC for each feature as a function of history length $N_{x}$. Again two features, autocorrelation $AC_{x,y}$ and assists $A_{x,y}$, are consistently accurate predictors across all values of $N_{x}$. For the autocorrelation feature, this robustness indicates that pairs of friends interact more periodically than non-friends, regardless of their overall level of activity in the system. This signal is strong despite common individual schedules (e.g., weekend nights) that could potentially lead to artificially high autocorrelation between non-friends. Furthermore, even when an individual's data is sparse because he or she has completed very few games (less than 10), both autocorrelation and direct assistance have surprisingly strong predictive power, yielding average AUC values close to 0.98.

Focusing on autocorrelation, the reason for its high accuracy at small history lengths $N_{x}$ is likely due to the large number of individuals in the system at any one time. This very large pool makes the probability very low for interacting with the same non-friend individual more than a few times. In real-world systems with low thresholds for two individuals meeting by chance (e.g., colocation in highly constrained or small physical environments), autocorrelation can be less discriminative and may require augmentation with other temporal or domain-specific features. Essentially, context can matter: it is unlikely that everyone who frequents the same busy coffee shop on Monday mornings will be friends, due to the nature of that location, while it would be a good bet that many pairs of individuals attending the same weekly soccer practice would be friends. The large effective capacity of an online system means that any signal from autocorrelation is likely to be significant.

In their analysis of friendship and gameplay in~{\em Reach}, Mason and Clauset showed that individuals who are friends tend to coordinate and cooperate in ways that increase their team's score and the probability of winning the match~\cite{mason2012friends}. The strongly predictive nature of direct assists $A_{x,y}$ that we observe corroborates this finding, and demonstrates that it holds over a wide range of $N_{x}$. That is, even for casual users, counting these prosocial interactions is a reliable indicator of friendship because friends do indeed cooperate more than non-friends. 

Autocorrelation and direct assistance both maintain high performance across all sizes of $N_{x}$. The temporal features of raw and normalized pair frequencies $N_{x,y}$ and $N_{x,y}/N_{x}$ are less reliable predictors for small histories, but become more reliable as $N_{x}$ increases. For large histories ($N_{x}>400$), both features reach AUC values of nearly 0.90. 

As we might have expected from our previous analysis, the performance of spatial and temporal entropy features $H_{s,t}(x)$, $H_{t}(x)$, and $H_{s}(x)$ do not improve as we accumulate more data. Similarly, we observe fairly weak improvements for indirect assists $A_{x,y}$ and betrayals $B_{x,y}$.

The remarkable accuracy achieved by our two best features, autocorrelation of schedules and direct assistance (prosocial interactions), demonstrate that lightweight predictors can be reliable even when applied to individuals with heterogeneous amounts of data by which to estimate latent friendships.

\section{Social network inference}
\label{sec:social:network}
Given the excellent performance and computational efficiency{\footnote{The autocorrelation function can be computed in $O(n\log n)$ time using a fast Fourier transform.}} of the autocorrelation of co-play feature, $AC_{x,y}$, we use this lightweight predictor of friendship to infer the social network of the entire population of 17 million players. For each pair of players in the interaction network we compute $AC_{x,y}$, compare it to a threshold, which we explain below, and then label the pair of players as friends if their $AC_{x,y}$ is greater than or equal to the threshold value.

\subsection{Threshold selection}
The survey respondents are a biased sample of \textit{Reach} players~\cite{mason2012friends}, being substantially more skilled than the typical player and investing roughly an order of magnitude more time playing than an average player. It is thus possible that the survey sampling bias has produced an oversampling or an undersampling of the tail of the degree distribution.
In an attempt to control these opposing biases, we choose two thresholds, one to show what the network looks like if the survey respondents have less friends (undersampled tail) than the population, and one to show network structure if the respondents have more (oversampled tail).\\

\begin{figure}[t!]
\begin{center}
\includegraphics[scale=0.45]{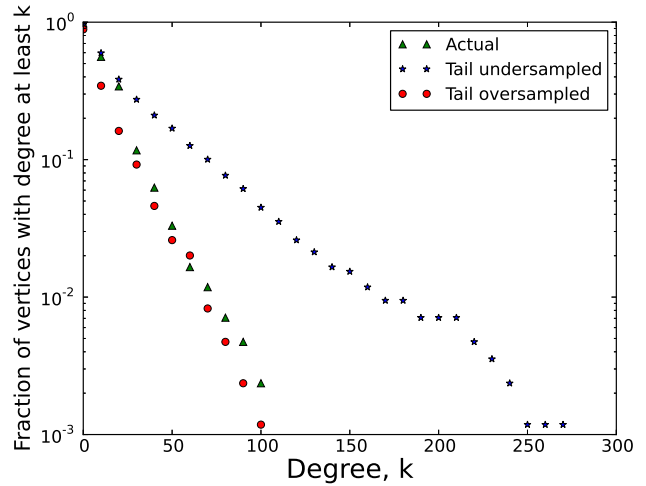}
\caption{CCDF of actual and inferred degree distributions using only survey respondent data.}
\label{fig:dd}
\end{center}
\end{figure}

\noindent{\bf Undersampled tail} - To control for the undersampled tail bias we choose the $AC_{x,y}$ that minimizes the Kullback-Leibler divergence
\begin{align}
D_{KL}(P||Q) = \sum_i \ln \left( \frac{P(i)}{Q(i)} \right)P(i) \enspace,
\end{align}
where $P$ is the degree distribution of social network derived from the survey respondent data and $Q$ is the degree distribution calculated by creating edges between players $x$ and $y$ if their $AC_{x,y}$ is greater than or equal to a chosen threshold. As shown in Figure~\ref{fig:dd}, this approach chooses $AC_{x,y}=197$ and produces an inferred degree distribution for the entire network of 17 million players that matches the density near the head of the actual distribution but with a heaver tail than the survey data. It is not clear that this threshold choice necessarily produces an abundance of false friendships, as players with many friends are unlikely to have reported them all due to the tedious and time consuming nature of providing this information via the survey. This hypothesis is supported by empirical research, which showed that self-survey respondents tend to underestimate their interactions with individuals as a function of recency~\cite{eagle2009inferring}. In our case, if a respondent did not interact with a friend recently, the tie may have been unreported.\\

\noindent{\bf Oversampled tail} - To control for the oversampled tail bias, we compute the threshold by finding largest $AC_{x,y}$ that produces a degree distribution with a maximum degree no larger than the maximum degree observed in the survey. This approach chooses $AC_{x,y}=1900$ and the tail of the inferred degree distribution agrees well with the survey data but less so near the head (see Figure~\ref{fig:dd}).

\begin{figure*}[t!]
\begin{center}
 \begin{tabular}{ccc}
 \includegraphics[scale=0.285]{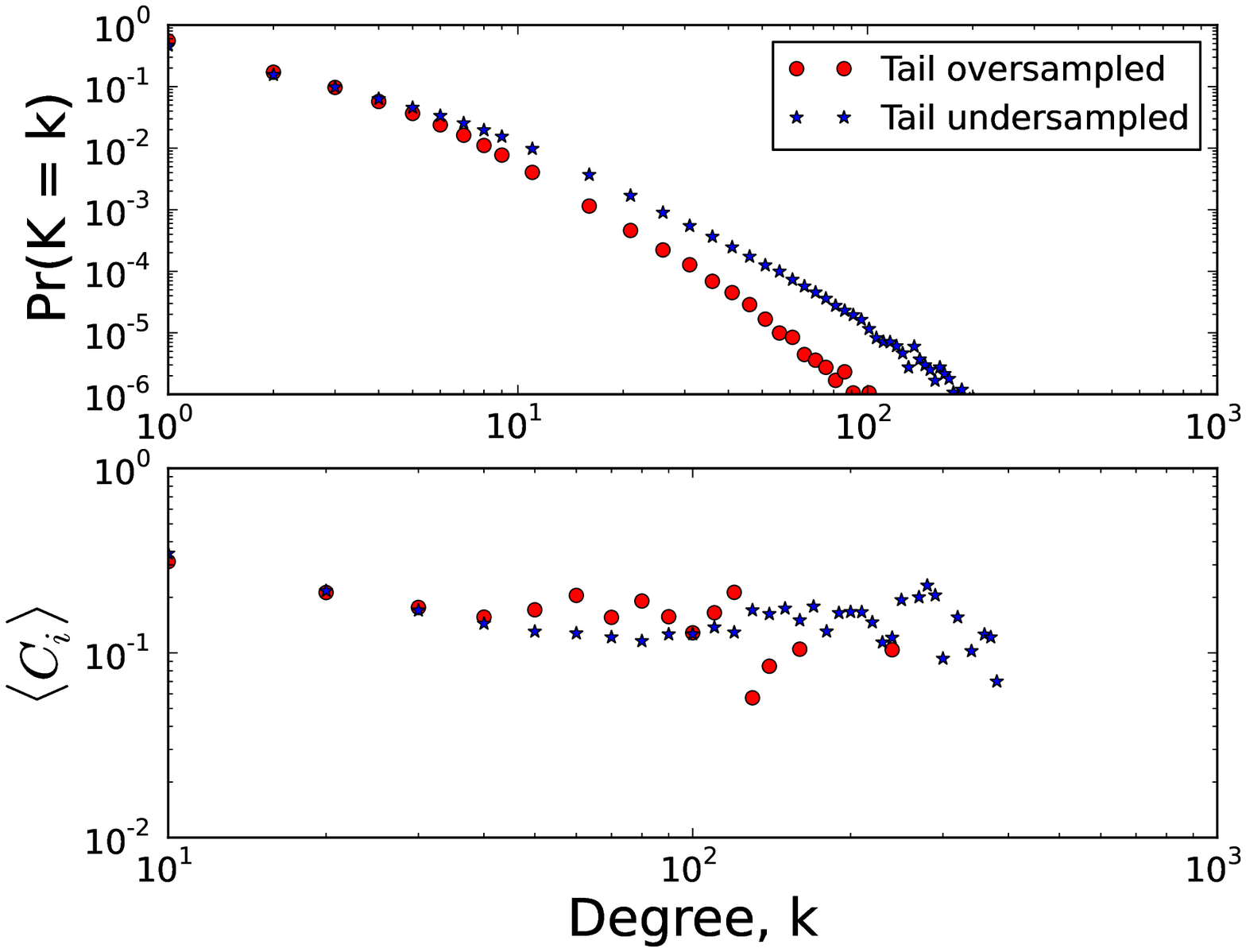}
 \includegraphics[scale=0.285]{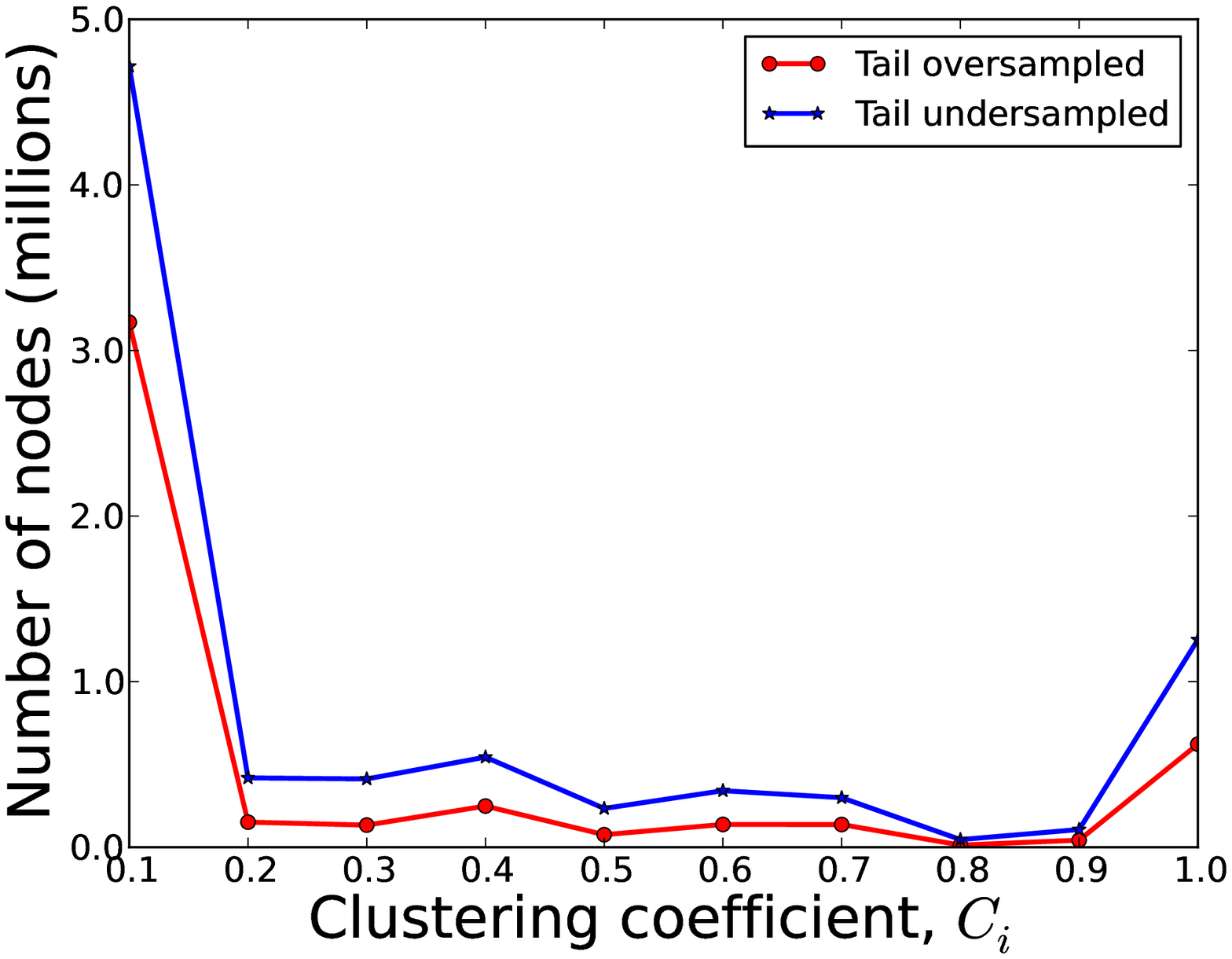}
 \includegraphics[scale=0.285]{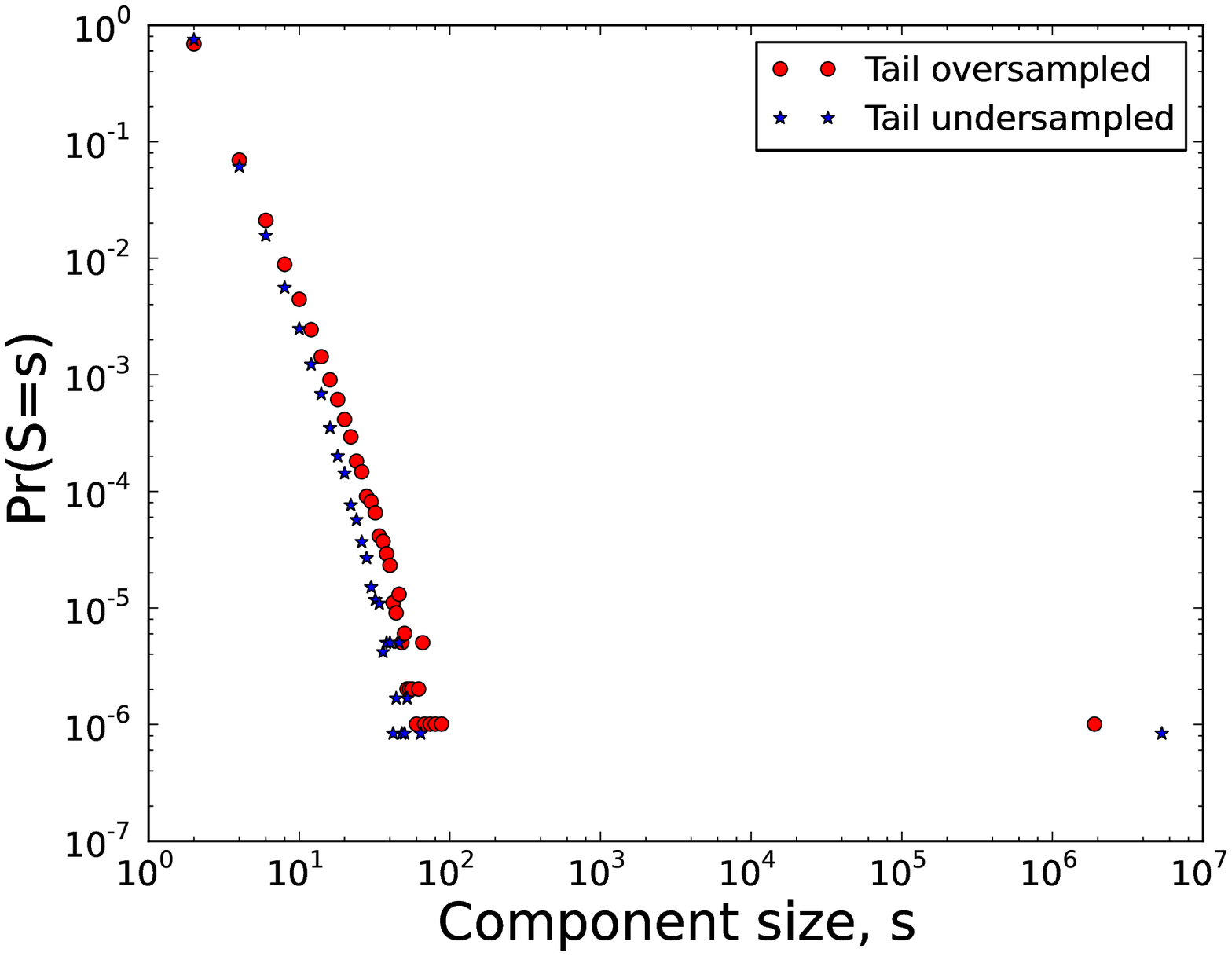}
 \end{tabular}
\caption{
(Left) Degree distribution and mean clustering coefficient, $\langle C_i \rangle$ as a function of degree for both thresholds using the entire population of players. (Center) Binned clustering coefficient, $C_i$, plots for both thresholds using the entire population of players, bin width $= 0.1$. (Right) Distribution of component sizes. The undersampled tail network contains 1,194,032 components. The oversampled tail network contains 991,932 components. 
} 
\label{fig:cc:hist:deg}
\end{center}
\end{figure*}

\subsection{Network structure}
These two thresholds represent reasonable bounds for what we expect for our interaction data as a whole. We now apply these two thresholds to the interactions among the full 17 million players and study the structure of the induced social network. In the undersampled tail scenario ($AC_{x,y}=197$), the inferred network consists of 8,373,201 nodes and 31,051,991 edges, while the network inferred using the oversampled tail threshold ($AC_{x,y}=1900$), contains 4,732,405 nodes and 11,435,351 edges.

The top panel of Figure~\ref{fig:cc:hist:deg}(Left) indicates that both cases we observe degree distributions with heavy tails, where the majority of nodes in the network are connected to a small number of neighbors while a small number of nodes are connected to a large number of neighbors. When compared to the social graph of Facebook discussed in~\cite{ugander2011anatomy}, players in \textit{Reach} have smaller numbers of friends. The median friend count in Facebook is 99 while in \textit{Reach} it is roughly $1/100^{th}$ the size, 1 and 2 at the over- and undersampled thresholds respectively. This large difference is likely caused by the high relative cost of establishing and maintaining a friendship in \textit{Reach} versus the more cost-free nature of Facebook friendships. Specifically, \textit{Reach} players must consistently and periodically interact over long periods of time, which is a significant investment of effort, while in Facebook, they must only click a request or accept button.

A vertex's clustering coefficient is defined as
\begin{align}
C_i = \frac{\textrm{number of connected neighbors}}{\textrm{number of possible connected neighbors}} \enspace ,
\end{align}
and provides a principled way of measuring how close vertex $i$ and its neighbors are to forming a clique~\cite{newman2010networks}. This statistic equals unity when a vertex and its neighbors form a clique, while it equals zero when none of its neighbors are themselves pairwise connected. In our inferred graph, shown in the bottom panel of Figure~\ref{fig:cc:hist:deg}(Left), a substantial fraction of individuals (between 16-20\%) form tightly knit groups with high values of $C_{i}$.

Furthermore, the functional relation between the mean clustering coefficient $\langle C_i \rangle$ as a function of degree $k_{i}$ is roughly the same, regardless of which threshold we choose (Fig.~\ref{fig:cc:hist:deg}(Center)). For example, even when a vertex has a degree of 100, its clustering coefficient is likely to be between 0.1 and 0.2. This suggests that threshold choice does not substantially change the underlying network structure, and these numbers are close to those estimated for the Facebook social graph, where the mean clustering coefficient for a vertex with degree 100 was 0.14~\cite{ugander2011anatomy}. While the mean clustering coefficient remains large independent of degree, a mild decreasing trend is evident. This suggests that nodes with high degree, who are likely high volume players, interact with others relatively less discriminately than nodes with smaller degrees, a pattern also found in the analysis of the Facebook social graph~\cite{ugander2011anatomy}. 

Figure~\ref{fig:cc:hist:deg}(Right) plots the distribution of component sizes and indicates that the network contains a single large connected component composed of between two and four million players. The majority of the remaining nodes are spread amongst many components containing between roughly ten and twenty nodes. In the case of an undersampled tail, the network contains 1,194,032 components. In the oversampled case, the network contains 991,932 components.


\section{Conclusion}
\label{sec:conclusion}
Our motivating question was whether latent social ties like friendships can be accurately recovered from interaction data alone, and indeed we have shown that they can, with remarkable accuracy. We demonstrated that periodicity between interactions and specific prosocial behaviors across these interactions are both highly robust indicators of friendship, even in instances where data are sparse. Information theoretic measures of spatial and temporal behavior, which are good indicators of the quantity of social ties in other contexts, are not effective at predicting the ties alone, but may be useful in combination with other temporal features. There are a number of interesting points these results suggest, both for improving \textit{Reach} and for enabling friendship-aware applications in other domains.

Many online games, including \textit{Halo: Reach}, rely on matchmaking algorithms to place individuals onto teams in order to make a new game instance go. If the \textit{Reach} matchmaking algorithm works as desired, the teams are equally matched and the competition's outcome is unpredictable. However, when individuals play with friends, their performance improves~\cite{mason2012friends}, and this synergy is not included in the calculations of the matchmaking algorithm. A friendship-aware matchmaking algorithm, using features like the ones we consider here, could correct for the effective increase in team skill that occurs when friends play together, without reference to an external ``friends list'', and thus produce better matched teams, more enjoyable gameplay and overall greater engagement by the users. Another improvement would be to suggest as friends (to be added to a user's friends list) those individuals with whom a player has exhibited significant prosocial interactions, such as direct assists.

In the more general context of an online system where we can observe interactions, but not labeled friendship ties, our results could be applied in an unsupervised manner. Using an unsupervised learning algorithm such as $k$-means to separate friends from non-friends based on the autocorrelation values of their co-interaction time series should be relatively simple and robust. The discriminatory power of autocorrelation and prosocial behavior, even with sparse data, suggests that latent friendship ties may in fact be easily detectable, due to the nature of friendship itself. In a sense, periodic and prosocial interactions are the definition of friendship, and it may be difficult to maintain such a relationship online without manifesting a signal in these ways.

Friendship-aware applications are only one new opportunity presented by the automatic inference of latent social ties from interaction data. The ease with which we were able to recover the latent friendship labels raises significant privacy questions, as these labels are often considered private information. The accurate recovery of such private signals from public interaction data may facilitate malicious applications. The social consequences of large-scale deployment of friendship inference is difficult to estimate.

Other benefits are more easily identified. For instance, many questions in computational social science may benefit from the accurate recovery of the underlying social network that generates the observed data. The general outlines of our results may have productive applications in many of these domains, e.g., in big data analyses of online social behavior. Our results are encouraging for settings where ground-truth data are at best rare and expensive to collect. Robust methods to extrapolate from ground-truth survey data to large-scale latent social network prediction are of great practical interest. We look forward to seeing the exploration of these and other beneficial applications.

\section{Acknowledgements}
We thank Christopher Aicher and Nora Connor for insightful comments and valuable feedback, Chris Schenk for his help developing the data acquisition system and web survey, and Bungie Inc.\ for providing access to the data. We acknowledge financial support from the James S.\ McDonnell Foundation.

\bibliography{refs}

\end{document}